\documentclass[prb,twocolumn,superscriptaddress,amssymb,nobibnotes]{revtex4-1}

\usepackage{xcolor}
\definecolor{tab20green}{rgb}{0.1725490196,0.6274509804,0.1725490196}
\definecolor{tab20red}{rgb}{0.8392156863,0.1529411765,0.1568627451}
\definecolor{tab20orange}{rgb}{1.0,0.4980392157,0.05490196078}
\definecolor{tab20blue}{rgb}{0.1215686275,0.4666666667,0.7058823529}
\usepackage{amsfonts,amssymb,amsmath}
\usepackage{graphicx}
\usepackage{dsfont}
\usepackage{soul}
\usepackage{ulem}
\usepackage[
  colorlinks,
  linkcolor=blue,
  urlcolor=blue,
  citecolor=blue,
  plainpages=false,
  pdfpagelabels,
]{hyperref}
\usepackage{pdfcomment}

\DeclareMathAlphabet\mathbfcal{OMS}{cmsy}{b}{n}

\newcommand{\baleqn}{\begin{equation}\begin{aligned}[b]}
\newcommand{\ealeqn}{\end{aligned}\end{equation}}
\newcommand{\baleqns}{\begin{equation*}\begin{aligned}}
\newcommand{\ealeqns}{\end{aligned}\end{equation*}}
\newcommand{\be}{\begin{equation}}
\newcommand{\ee}{\end{equation}}

\newcommand{\beq}{\begin{equation}}
\newcommand{\eeq}{\end{equation}}
\newcommand{\beqa}{\begin{eqnarray}}
\newcommand{\eeqa}{\end{eqnarray}}

\begin{document}
\title{Topological bulk states and their currents}

\author{Chris N. Self}
\thanks{CNS and AR-G contributed equally to this work.}
\affiliation{School of Physics and Astronomy, University of Leeds, Leeds LS2 9JT, UK}
\author{Alvaro Rubio-Garc\'{i}a}
\thanks{CNS and AR-G contributed equally to this work.}
\affiliation{Instituto de Estructura de la Materia IEM-CSIC, Calle Serrano 123, Madrid E-28006, Spain}
\author{Juan Jose Garc\'{i}a-Ripoll}
\affiliation{Instituto de F\'{i}sica Fundamental IFF-CSIC, Calle Serrano 113b, Madrid E-28006, Spain}
\author{Jiannis K. Pachos}
\affiliation{School of Physics and Astronomy, University of Leeds, Leeds LS2 9JT, UK}

\date{\today}
\pacs{...}

\begin{abstract}
We provide evidence that, alongside topologically protected edge states, two-dimensional Chern insulators also support localised bulk states deep in their valance and conduction bands. These states manifest when local potential gradients are applied to the bulk, while all parts of the system remain adiabatically connected to the same phase. In turn, the bulk states produce bulk current transverse to the strain. This occurs even when the potential is always below the energy gap, where one expects only edge currents to appear. Bulk currents are topologically protected and behave like edge currents under external influence, such as temperature or local disorder. Detecting topologically resilient bulk currents offers a direct means to probe the localised bulk states.

\end{abstract}

\maketitle

\section{Introduction}

Since their discovery topological materials have been very well studied theoretically~\cite{kane2005quantum, kane2005z2topological, bernevig2006quantum, hasan2010colloquium, qi2011topological,vergniory2019complete, tang2019topological, tang2019comprehensive}. More recently, however, the study has moved beyond theory. There is now a great deal of work focused on finding real materials with topological characteristics \cite{zhang2009topological, hsieh2008topological, chen2009experimental}. If we were able to harness topological materials for technological applications the resilience of edge states would make them extremely attractive for a number of applications: from frictionless directed transport of currents; to transistors, amplifiers and detectors \cite{mondal2010tuning,zhang2015electrical, banerjee2014topological, scharf2016tunneling, tanaka2009manipulation}. Though many of the topological materials discovered so far are unsuitable for use in technologies for reasons such as their toxicity, it is believed that about 27\% of recorded inorganic crystals have a topological nature\ \cite{vergniory2019complete} so maybe one day a material will be found that meets all the criteria for a practically useful topological material.

Among topological materials, Chern Insulators (CI's) are of much interest due to their theoretical simplicity~\cite{bernevig2013topological} and the fact that they can be realised in the laboratory with cold-atom technology~\cite{tarruell2012creating, Jotzu2014, Goldman2016, Flaschner2016, Asteria2019}. When two-dimensional CI's have sharp boundaries they develop gapless edge states~\cite{kane2005quantum, kane2005z2topological, bernevig2006quantum, hasan2010colloquium, qi2011topological}. These edge states support edge currents that have a number of remarkable properties. They flow without losses, flowing freely around imperfections on the edge with no backscattering. Further, their conductivity is quantised by the Chern number, which is a topological invariant that characterises the system. The common view of CI's is that it is these edge states that make them interesting, while the bulk is insulating and thus lacks any physical manifestation of its topological behaviour.
In this paper we challenge the idea that CI's are composed of robust edge states surrounding an inert bulk. We focus on the bulk of the CI and, by studying the influence of locally varying potentials, we find and characterise currents in the bulk of the system. These currents arise from a continuous distortion of the valence band that gives rise to localised bulk states. To understand their origin one can view the bulk of a homogeneous Chern insulator as a perfectly entangled state of pairs of edge-like currents, adding up to zero net flow. Potential gradients strain those states, progressively disentangling the hidden currents through a transfer of population. The bulk states share the robustness of edge states. Beyond this, they have an enhanced tuneability and no geometric constraints imposed by the shape of the system. Hence, in addition to being interesting, they potentially point the way to another approach for designing technologies based on topological materials, which will perhaps be easier to engineer.

The paper is organised as follows. In Section~\ref{sec:Edge-and-bulk-currents} we recap the physics of the edge of a CI and we derive the topological edge current in terms of the Chern number $\nu$. We also summarise our main findings concerning bulk currents in CI's. In Section~\ref{sec:currents-in-lattice-models} we consider a concrete latice model of a CI, the Haldane model\ \cite{haldane1988model}, and we show the presence of topological localised edge and bulk currents. In particular, we demonstrate the robustness and topological character of the bulk currents. In Section~\ref{sec:Origin-of-bulk-currents} we perturbatively calculate the effect a potential gradient has in 1D and quasi-2D topological systems. We demonstrate that such potential gradients result in localised modes in the bulk, which are crucial for the appearance of localised bulk currents. Finally we sum up in Section~\ref{sec:Conclusions}.

\section{Edge and bulk currents}
\label{sec:Edge-and-bulk-currents}

We first determine the currents supported by the localised edge states of a CI by considering it in the continuum. We model the edge physics of a CI with an isolated, one-dimensional, fermionic Hamiltonian $H = \nu\int_{-\infty}^\infty dp \, \varepsilon(p) \, a^\dagger_pa_p$\ \cite{bernevig2013topological}. The annihilation and creation operators for momentum $p,$ $a^\dagger_p$ and $a_p,$ obey a linear dispersion relation, $\varepsilon(p) = p$, that extends to infinity. In equilibrium at temperature $T,$ the occupation of the edge states depends on momentum and chemical potential $\mu$ as $n(p) = (e^{(\varepsilon(p) - \mu)/T}+1)^{-1}$. Holes in the negative energy modes are quasiparticles with positive energy and momentum, which we write as $b_{p}^\dagger = a_{-p}.$ In units of $e^2/\hbar,$ the net particle current is
\begin{equation}
I_\text{edge} = \frac{\nu}{2\pi} \int_0^\infty d\varepsilon \left[n(\varepsilon-\mu)-n(\varepsilon+\mu)\right].
\label{eqn:preden}
\end{equation}
Evaluating this integral we find that a uniform chemical potential $V(\boldsymbol{r})=-\mu$ acting on a material with Chern number $\nu,$ induces a current
\begin{equation}
\label{eq:current1}
    I_\text{edge} = \frac{\nu}{2\pi} \, \mu,
\end{equation}
which is robust against temperatures and local disorder.

The main result of this paper is that, surprisingly, if the chemical potential is non-uniform and varies over the system it \textit{creates topological currents in the bulk of the material}. Specifically, any small potential gradient $\boldsymbol{\nabla} V(\boldsymbol{r})$ induces a local perpendicular bulk current
\begin{equation}
\label{eq:current2}
    I_\text{bulk} (\boldsymbol{r}) = \frac{\nu}{2\pi}\,a_0|\boldsymbol{\nabla} V(\boldsymbol{r})|,
\end{equation}
proportional to the Chern number $\nu$ and the lattice constant $a_0$. This can occur without the presence of gapless states in the bulk. It is observed even when the potential is so small that it cannot close the insulating gap, cause a phase transition or locally destroy topological order. Hence, these currents emerge not due to the population of new conducting orbitals above the Fermi level, but because of a particular restructuring of the valance band states that give rise to \textit{localised bulk states}, as we shall see in the following. Furthermore, these bulk currents share the topological protection of edge currents against finite temperature as well as local disorder. Unlike edge currents, the bulk currents have a direction and geometry that are tuneable with the potential $V(\boldsymbol{r}),$ unconstrained by the form of the sample.

For an integer quantum Hall system with Chern number $\nu$ we can investigate the presence of bulk currents through their description in terms of a Chern-Simons theory. In the absence of a boundary the action that describes the system is given by
\be
S[A] = {\nu\over 4\pi} \int d^2x dt \epsilon^{\mu\kappa\lambda} A_\mu\partial_\kappa A_\lambda,
\ee
where $A$ is an Abelian vector potential. The electric current density\ \cite{nissinen2018qhe, maeda1996chiral} is given by
\begin{equation}
I^\mu_{\rm{bulk}} = {\delta S[A] \over \delta A_\mu} = {\nu \over 2\pi} \epsilon^{\mu\kappa\lambda} \partial_\kappa A_\lambda.
\label{eqnIQHE}
\end{equation}
When $\mu=x$, $\kappa=y$ and $\lambda=t$ we obtain the continuous version of formula \eqref{eq:current2}, where the current ${\boldsymbol I}_{\rm{bulk}}$ is perpendicular to the gradient $\boldsymbol{\nabla} A_0$ and $A_0=V$ is the local potential. In the following we shall investigate the properties of bulk currents in a microscopic system and analyse the physical reason that gives rise to their presence.

\section{Currents in lattice models}
\label{sec:currents-in-lattice-models}

In this Section we investigate the emergence of topological edge and bulk currents in the Haldane model, a 2D Chern insulator defined on a hexagonal lattice. This allows us to verify the validity of Eqn.\ \eqref{eq:current2} and probe the stability of bulk currents in the presence of finite temperature and disorder.

\subsection{Lattice currents}
\label{sec:fourier}

Consider a lattice model of free fermions $\{c_i,\, c^\dagger_i\},$ with a quadratic Hamiltonian
\begin{equation}
H = \sum_{i,j} A_{ij} \, c^\dagger_i c_j \,.
\label{eqn:general-hopping-model}
\end{equation}
The Hermitian matrix $A_{ij}$ contains both hopping terms $i \neq j$ as well as on-site potentials $i = j$. We particularize to systems with a cylindrical symmetry. The periodic coordinate is labelled $x,$ the height on the cylinder is $y$, and both coordinates have finite lengths $L_x$ and $L_y$.

We regard the cylindrical problem as a one-dimensional model with translational invariance along the $x$ coordinate, and a large one-dimensional \textit{``unit cell"} of size $L_y$. We Fourier transform the fermionic operators along the periodic coordinate $x,$ introducing the momentum coordinate $p$
\begin{equation}
c^\dagger_{j,p} = \frac{1}{\sqrt{L_x}} \sum_x e^{-ip x} c^\dagger_{j,x} \,.
\end{equation}
The momentum takes discrete values $p = 2 \pi n / L_x$ for $n = 1,2,\ldots,L_x,$ and the index $j\in\{1,2,\ldots L_y\}$ labels the position in the unit cell. The Hamiltonian separates into a sum of contributions $H(p),$ one for each momentum $p$
\begin{equation}
H = \sum_p H(p) = \sum_{p} \sum_{j,l} A_{jl}(p) \, c^\dagger_{j,p} c_{l,p} \, .
\end{equation}
The diagonalization of matrix $A_{jl}(p)\in\mathbb{C}^{L_y\times L_y}$ with a unitary matrix $U_{jm}(p)$ provides us with the fermionic eigenmodes of $H(p)$,
\begin{equation}
d^\dagger_{m,p} = \sum_j U_{jm}(p) \, c^\dagger_{j,p} \,.
\end{equation}
The correlator between any two points in real space, $(x,y)$ and $(r,s),$ is a function of the correlations in momentum space. In thermal states, these mode occupations will follow a Fermi-Dirac statistics, $\langle d^\dagger_{m,p} d_{m',p'} \rangle = n_T\left(\varepsilon_{m}(p)\right)\delta_{m,m'}\delta_{p,p'}$ and we may write
\begin{equation}
\langle c^\dagger_{xy} c_{rs} \rangle = \frac{1}{L_x} \sum_{m p} e^{-ip(x-r)} \, U^*_{ym}(p) U_{sm}(p) \, n_{T,\mu}\big(\varepsilon_{m}(p)\big).
\label{eq:correlations}
\end{equation}
with an implicit dependence on the temperature and the chemical potential, via the Fermi function $n_{T,\mu}(\epsilon).$

The continuity equation for the site occupancy operator $n_i = c_i^\dagger c_i$ provides us with a definition of the particle current $J_{ij}$ flowing from $i$ to $j$ as $\textrm{d}n_i/\textrm{d}t = \sum_{j} J_{ij}.$ Through the Heisenberg equation $\textrm{d}n_i/\textrm{d}t = -i[n_i,H]$, this expression particularizes for quadratic Hamiltonians to
\begin{equation}
\sum_j J_{ij} = -i \sum_{m,n} A_{mn} \left[ c_i^\dagger c_i, c_m^\dagger c_n \right] .
\end{equation}
Given that the commutator evaluates to $[ c_i^\dagger c_i, c_m^\dagger c_n ] = \delta_{im} c_i^\dagger c_n - \delta_{in} c_m^\dagger c_i$, we identify the current operator as
\begin{equation}
J_{ij} = -i \left( A_{ij} \, c_i^\dagger c_j - A_{ji} \, c_j^\dagger c_i  \right).
\end{equation}
Taking expectation values provides us the average current between any two neighboring sites of the lattice
\begin{equation}
\langle {J}_{ij} \rangle = 2\ \textrm{Im}\left\lbrace A_{ij} \langle c^\dagger_i c_j\rangle\right\rbrace \, .
\label{eq:two_point_currents}
\end{equation}
We will need to study currents that flow over extended regions, such as the edge or the bulk of the system. In these cases we first define a boundary or line $l$ and compute the average current by summing all the $J_{ij}$ that start in a given site $i$ and cross that line.

\subsection{Haldane model on a cylinder}

\begin{figure}[t]
\centering
\includegraphics[width=\columnwidth]{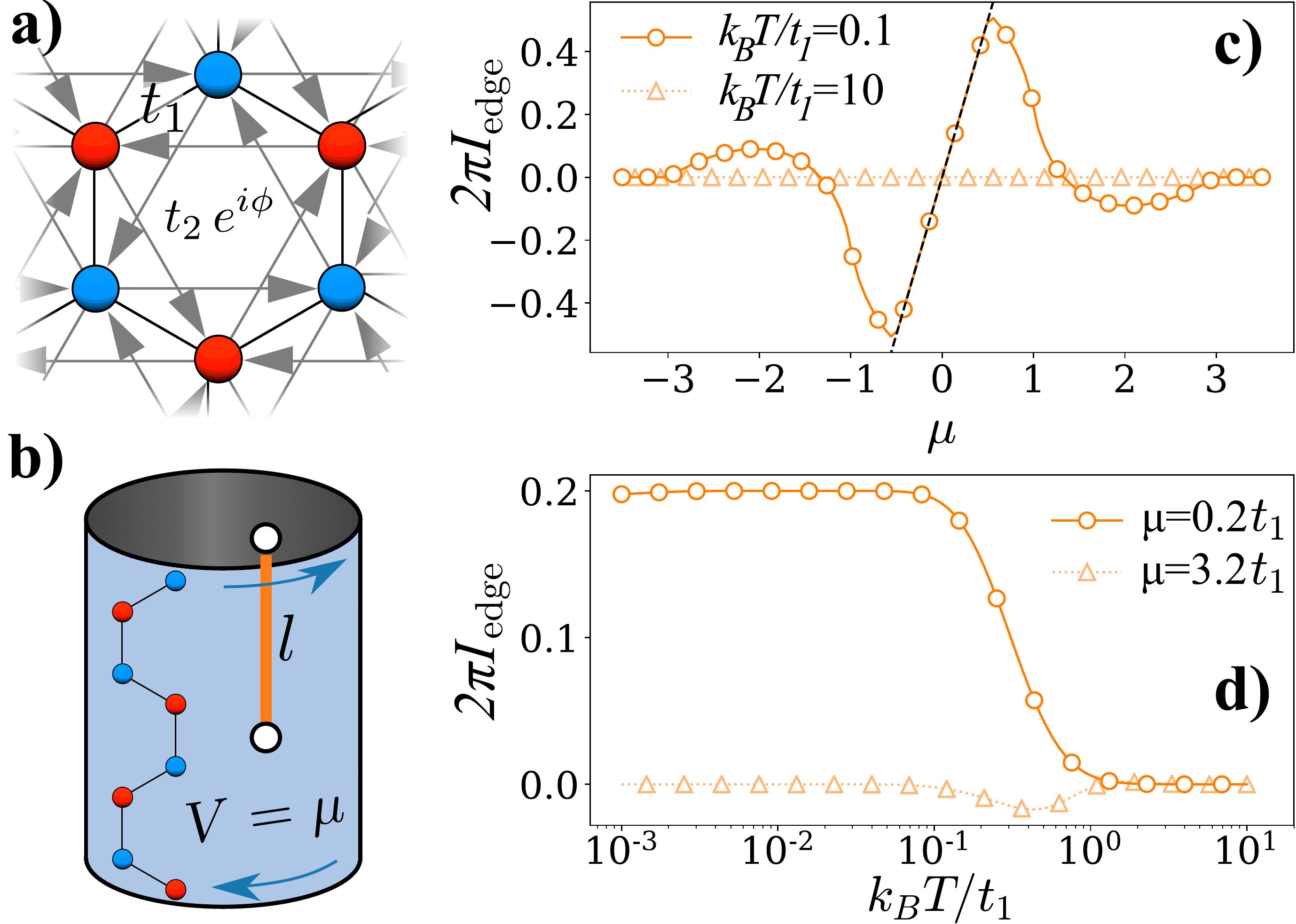}
\caption[current]{Edge currents of the Haldane model as a function of temperature, $T$, and chemical potential, $\mu$. (a) Hexagonal plaquette of the Haldane lattice model. Arrows denote hopping directions for which $\nu_{ij}=+1$ in Eqn.~\eqref{eqn:haldane-model-ham}. (b) The Haldane model on a cylinder showing the direction of the edge currents and the line $l$ used to define $I_\text{edge}$. (c) Edge currents vs. chemical potential for $k_BT=0.1t_1$ and $10t_1$. At low temperatures the current is independent of temperature and follows Eqn. \eqref{eq:current1}; at large temperatures it tends to zero. (d) Edge current vs. temperature, for $\mu=0.2t_1$ and $3.2t_1.$ When $\mu$ is small the edge current is stable against temperature. These results are obtained for a cylinder with $L_x\times L_y=300\times 30.$}
\label{fig:global-mu-currents}
\end{figure}

Haldane's Chern insulator~\cite{haldane1988model} is a fermionic hopping model of the form Eqn.~\eqref{eqn:general-hopping-model} defined on a honeycomb lattice with Hamiltonian
\begin{align}
H &= \sum_{\langle ij\rangle} t_1 \, c^\dagger_i c_j + \sum_{\langle\langle ij\rangle\rangle} t_2 \, \textrm{e}^{i\nu_{ij}\phi} c^\dagger_i c_j + \sum_i V(\boldsymbol{r}_i) \, c^\dagger_i c_i.
\label{eqn:haldane-model-ham}
\end{align}
Fig.\ \ref{fig:global-mu-currents}(a) sketches the hexagonal plaquette of the model, illustrating the real-valued nearest neighbour hopping $t_1$ and the complex-valued next-nearest neighbour hopping $t_2$ that has direction $\nu_{ij}=\pm 1$ and phase $\phi.$

The Haldane model has two gapped topological phases with Chern numbers $\nu = \pm1$, as well as a trivial phase where $\nu =0$. It is possible to cross between these phases by tuning $\phi$, $t_1$, and $t_2.$ Without loss of generality, we show numerical results for the Chern insulator phase with $\nu=+1,$ using $t_1 = 1$, $t_2 = 0.1,$ $\phi = \pi/2$.

The Haldane model is defined on a bipartite lattice, with two triangular sublattices A and B, colored red and blue in Fig.~\ref{fig:global-mu-currents}(a). It is convenient to write Hamiltonian~\eqref{eqn:haldane-model-ham} using two sets of fermionic operators $\{a_i,\, a^\dagger_i\}$ and $\{b_i,\, b^\dagger_i\},$ denoting sites in the $A$ and $B$ sublattices. As specified in Sect.\ \ref{sec:fourier}, we embed our problem in a cylinder and perform a Fourier transform along the periodic coordinate $x$. The Hamiltonian for fixed momenta $H(p)$ adopts the form
\begin{equation}
H(p) = \psi_{j,p}^\dagger K_1(p) \psi_{j,p} + \psi_{j,p}^\dagger K_2(p) \psi_{j+1,p} + \text{h.c.}
\label{eq:fourier_modes}
\end{equation}
with 
\begin{equation}
K_1(p) = {1 \over 2} \begin{pmatrix}
4t_2 \cos (\phi +p) +V_a(j) & t_1(e^{ip} +1)\\
 t_1(e^{-ip} +1) & 4t_2 \cos (\phi -p) +V_b(j)
\end{pmatrix}
\end{equation}
\begin{equation}
K_2(p) = \begin{pmatrix}
t_2 (e^{i\phi} +e^{-i(\phi-p)}) & 0\\
t_1 & t_2 (e^{-i\phi} +e^{i(\phi+p)})
\end{pmatrix}
\end{equation}
and $\psi_{j,p} = (a_{j,p} \,\,b_{j,p})^T$.


Since the unit cell of the honeycomb lattice contains $2$ sites (one in each sublattice), a one-dimensional cut of the lattice [cf.\ Fig.~\ref{fig:global-mu-currents}(b)] gives us $2 L_y$ sites and $2L_y$ different eigenvalues of $H(p),$  $\varepsilon_m(p)$ for $m = 1,2,\ldots,2 L_y.$ We refer to these as the `bands' of the model.

In the topological phase $\nu\neq 0,$ the cylinder supports two sets of edge states, with a gapless and approximately linear dispersion relation, $\omega\propto p.$ There are two edge modes for each $p,$ each of them localised at a different boundary of the cylinder and with group velocity in opposite directions, as sketched in Fig.~\ref{fig:global-mu-currents}(b).

\subsection{Edge currents}

To study the edge currents we specify a line transverse to the boundary of the system `$l$' [cf. Fig.\ \ref{fig:global-mu-currents}(b)] and analyze the total particle current across that line. The net particle current from `left' $(L)$ to the `right' $(R)$ on the top edge is a sum of current operators $I_{\rm edge} = \sum_{i\in L,j\in R} \langle {J}_{ij} \rangle,$ connecting vertices $i$ and $j$ on both sides of the line. Using Eqs.\ \eqref{eq:correlations} and \eqref{eq:two_point_currents} we have computed these currents for thermal states with temperature $T$ and chemical potential $\mu.$

Figs.~\ref{fig:global-mu-currents}(c-d) show the evolution of the edge currents $I_\text{edge}$ as we vary the temperature $T$ and the chemical potential $\mu.$ At low temperatures $k_B T\ll t_1,$ the edge current is proportional to the chemical potential $V(\boldsymbol{r})=\mu$ and satisfies $I_\text{edge} = \nu \mu/(2\pi)$, agreeing with Eqn.~\eqref{eq:current1} in Section~\ref{sec:Edge-and-bulk-currents}. We also observe that the current is invariant over a broad range of temperatures $T$, as expected~\cite{leonforte2019haldane}. Finally, the edge current vanishes when the temperature or the chemical potential approach the insulator gap. This is caused by the creation of particles and hole excitations in the conductance and valence bands.

\subsection{Bulk currents}
\label{sec:Bulk-currents-in-lattice-models}

\begin{figure}[t]
\centering
\includegraphics[width=\columnwidth]{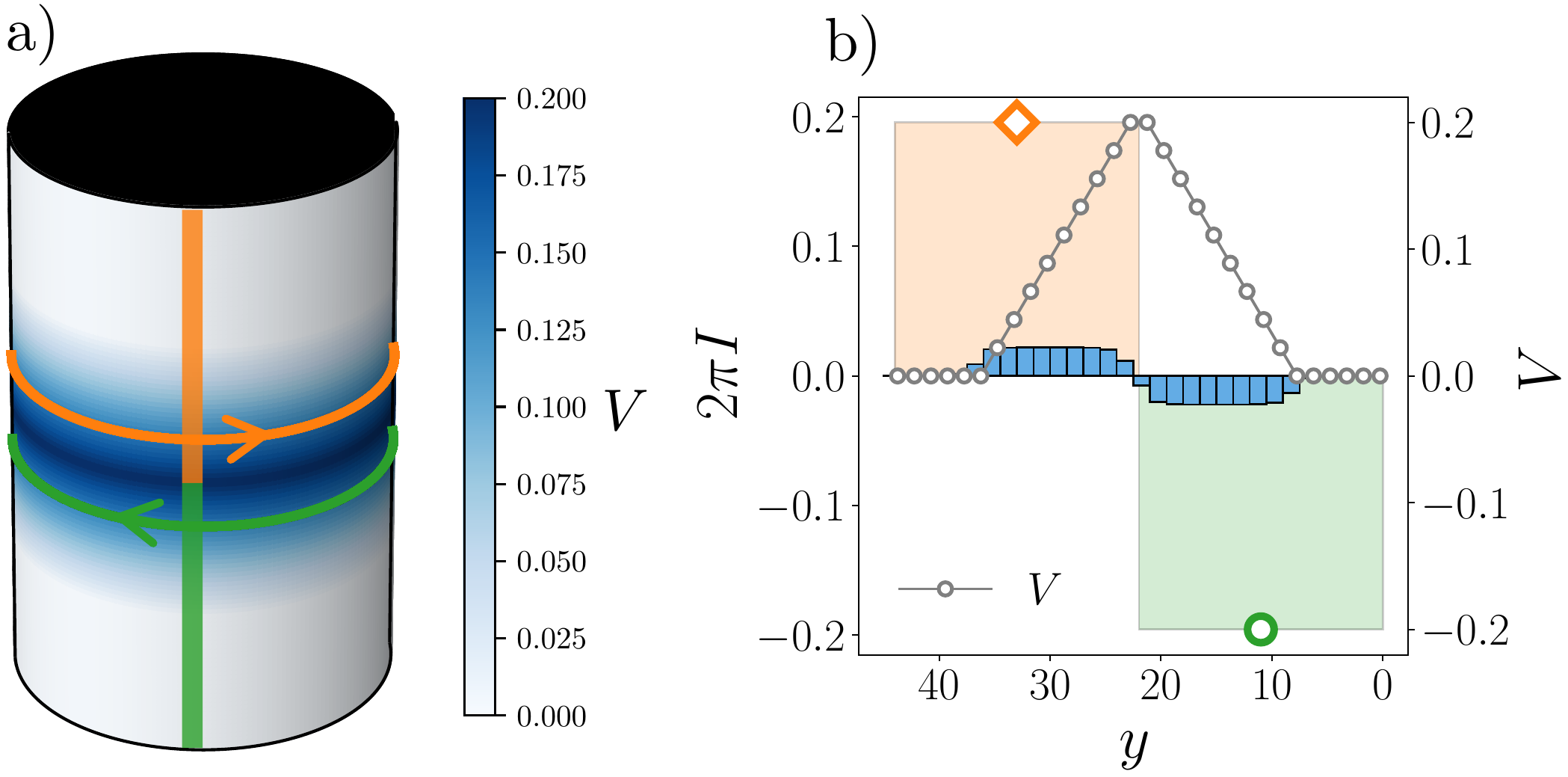}
\caption[current]{Spatial variations of the local potential give rise to bulk currents. The local potential is kept always below the energy gap, so the whole cylinder is always in the same topological phase. (a) Within a central band the potential linearly increases and decreases, with its maximum at the centre of the cylinder. (b) The currents $I$ as a function of height on the cylinder $y$. We observe bulk currents (blue bars) traverse to the $y$ axis around the whole cylinder with local strength that depends on the magnitude and sign of the potential gradient ${\boldsymbol{\nabla}} V$, in agreement with Eqn.~\eqref{eq:current2}. We also compute the total current along the top (orange diamond marker) and bottom (green circle) halves of the cylinder, finding that they are equal to the total change of potential along those regions. Data is presented for a cylinder with $L_x\times L_y=1000\times 30$ at\ $T=0$.}
\label{fig:triangle-potn-currents}
\end{figure}

Let us now investigate how the bulk of a Chern insulators reacts to inhomogeneous potentials, and in particular how new currents appear with a local potential that changes along the cylinder $V(y).$ The investigation is limited to potential strengths that are weaker than the insulator's energy gap, which means that $V(y)$ cannot induce a topological phase transition at any point on the cylinder. Moreover, we set the potential to zero at the boundaries $V(0)=V(L_y)=0,$ so that according to \eqref{eq:current1} no edge currents are excited $I_\text{edge}=0$ and we can focus strictly on the bulk physics.

Our first example is a potential that forms a triangular ramp along the cylinder, as shown in Fig.~\ref{fig:triangle-potn-currents}. This potential has uniform gradient ${\boldsymbol{\nabla}} V$ along certain bands around the cylinder and changes sign at the peak.  In the bands where the gradient is not zero, we observe currents that flow orthogonally to the gradient, with constant strength and a sign that depends on the sign of the gradient [cf. Fig.~\ref{fig:triangle-potn-currents}(b)].  In contrast, the regions where the potential is constant, exhibit no net current.

\begin{figure}[t]
\centering
\includegraphics[width=\columnwidth]{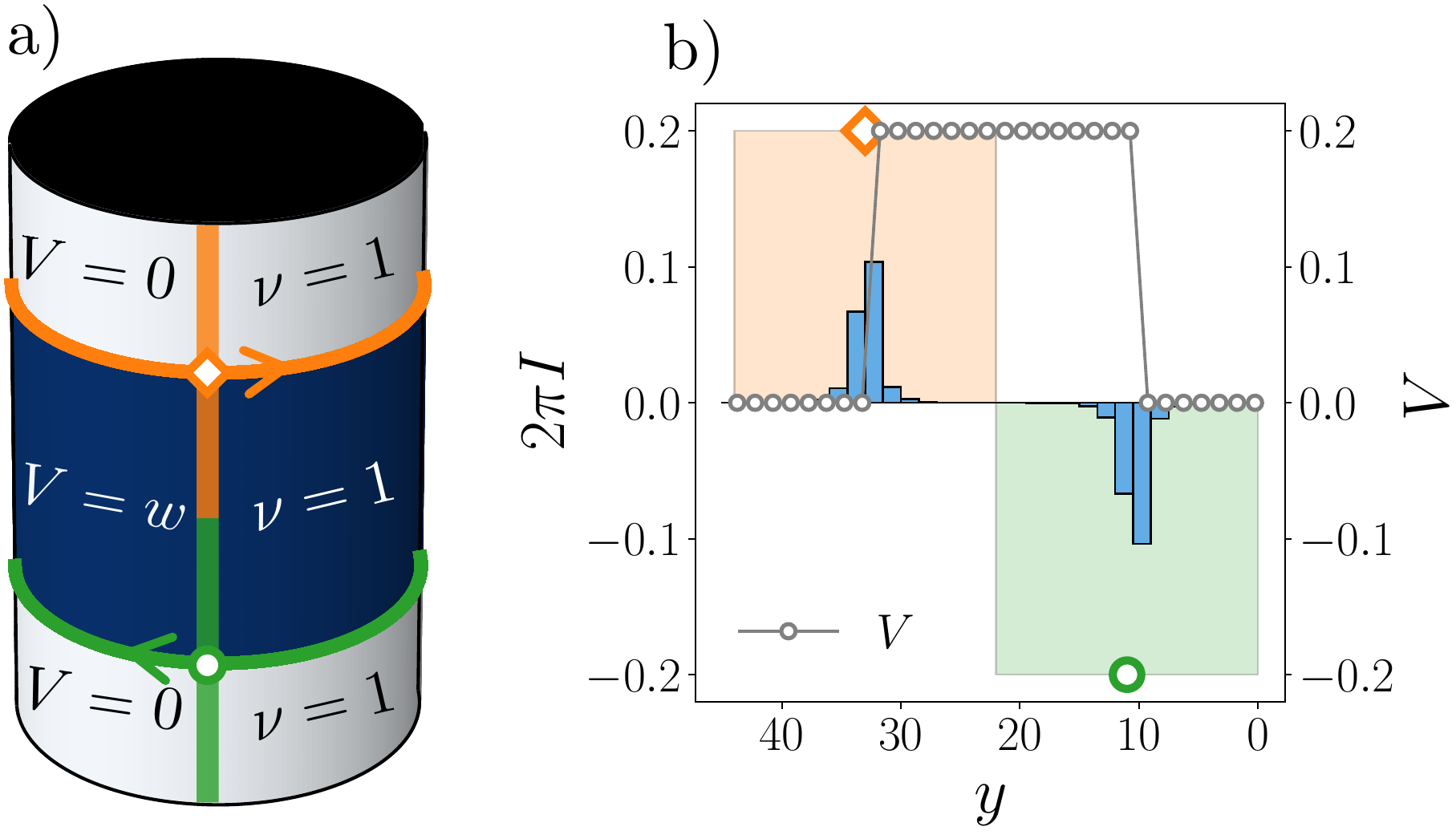}
\caption[current]{Bulk currents localised on potential steps at $y=L_y/4$ and $y=3L_y/4$. This is achieved by a stripe of potential around the cylinder. The currents are zero everywhere except at the potential steps, up to a regularisation by the lattice spacing. Data is presented for a cylinder with $L_x\times L_y=1000\times 30$ at\ $T=0$.}
\label{fig:middle-step-potn-currents}
\end{figure}

The previous example suggests that the CI reacts to a potential gradient by developing bulk currents. We will confirm this idea through a quantitative study of these currents. To do this we change the potential from a ramp to a double step function
\begin{equation}
    V(\boldsymbol{r}) = \left\{\begin{array}{ll}
         +w,& L_y/4\leq y \leq 3L_y/4, \\
         0,& \mathrm{else} \, .
    \end{array}\right.
    \label{eq:stripe}
\end{equation}
This function creates two potential steps at $y = L_y/4$ and $y = 3L_y/4$ as cartooned in Fig.~\ref{fig:middle-step-potn-currents}(a). From Eqn.~\eqref{eq:current2} we would expect a potential step to produce a delta function current localised on the jump (regularised in the simulations by the lattice spacing), with a strength proportional to the height of the step $w$.  A numerical study confirms this prediction, showing two bulk currents narrowly localised around the stripe boundaries, $y = L_y/4$ and $y = 3L_y/4$. These two currents have the same strength, but flow along opposite directions, as shown in Fig.~\ref{fig:middle-step-potn-currents}(b).

\begin{figure}[t]
	\centering
	\includegraphics[width=\columnwidth]{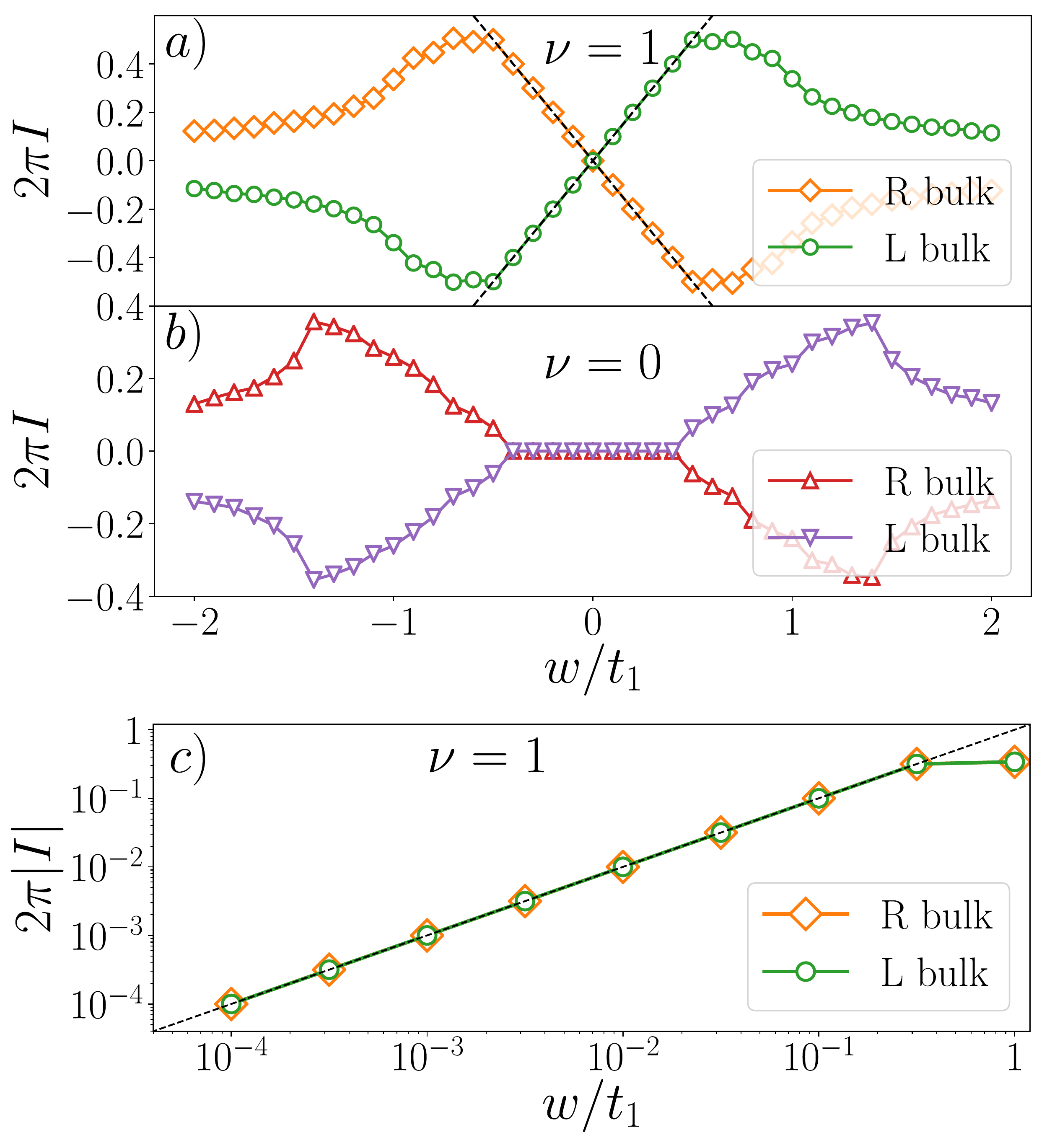}
	\caption[current]{Systematic investigation of the dependence of the bulk currents on the topological properties of the system. We study a cylinder with a central stripe of increased potential (illustrated in Fig.~\ref{fig:middle-step-potn-currents}(a)). (a) Bulk currents scale linearly with the potential gradient when the system is in a topological phase\ $M=0,\ \nu = \pm 1$. (b) Outside the topological phase,\ $M=0.9\,t_1,\ \nu=0$, the bulk currents vanish in the gapless region, consistent with Eqn.\ \eqref{eq:current2}. (c) In a phase with nonzero Chern number,\ $M=0$, localized bulk currents appear even for exponentially small steps\ $w/t_1$. Computations were carried out on a system of size $L_x\times L_y=1000\times 30$ at\ $T=0$.}
	\label{fig:topological-currents}
\end{figure}

In Fig.~\ref{fig:topological-currents}(a) we investigate quantitatively the relationship between the strength and sign of the localized currents $I$ and the value of the potential jump $w$. For small $w$, currents scale linearly with $w$ as would be expected from Eqn.~\eqref{eq:current2}. We suspect that, similar to edge currents, the bulk currents are direct manifestations of the non-trivial topological properties of the model. If this is true, the currents should depend on the Chern number $\nu$ of the system and vanish when we cross into the non-topological phase. To confirm this we introduce a dimerisation potential\ \cite{haldane1988model}
\begin{equation}
V_d = M \sum_i \left( a^\dagger_i a_i - b^\dagger_j b_j\right)
\label{eq:dimerisation_potential}
\end{equation}
that breaks the topological phase for\ $|M/t_2| > 3\sqrt{3}|\sin \phi|$. As shown in Fig.\ \ref{fig:topological-currents}(b), we indeed find that bulk currents disappear when we enter the\ $\nu=0$ phase with a dimerisation potential\ $M=0.9\,t_1$. In both the trivial insulator and topologcial phases greater values of $w$ close the band gap and allow trivial, non-topological currents. Finally, Fig.~\ref{fig:topological-currents}(c) confirms the linear growth of the bulk currents with the potential gradient, even for very small values of $w.$

\begin{figure}[t]
	\centering
	\includegraphics[width=\columnwidth]{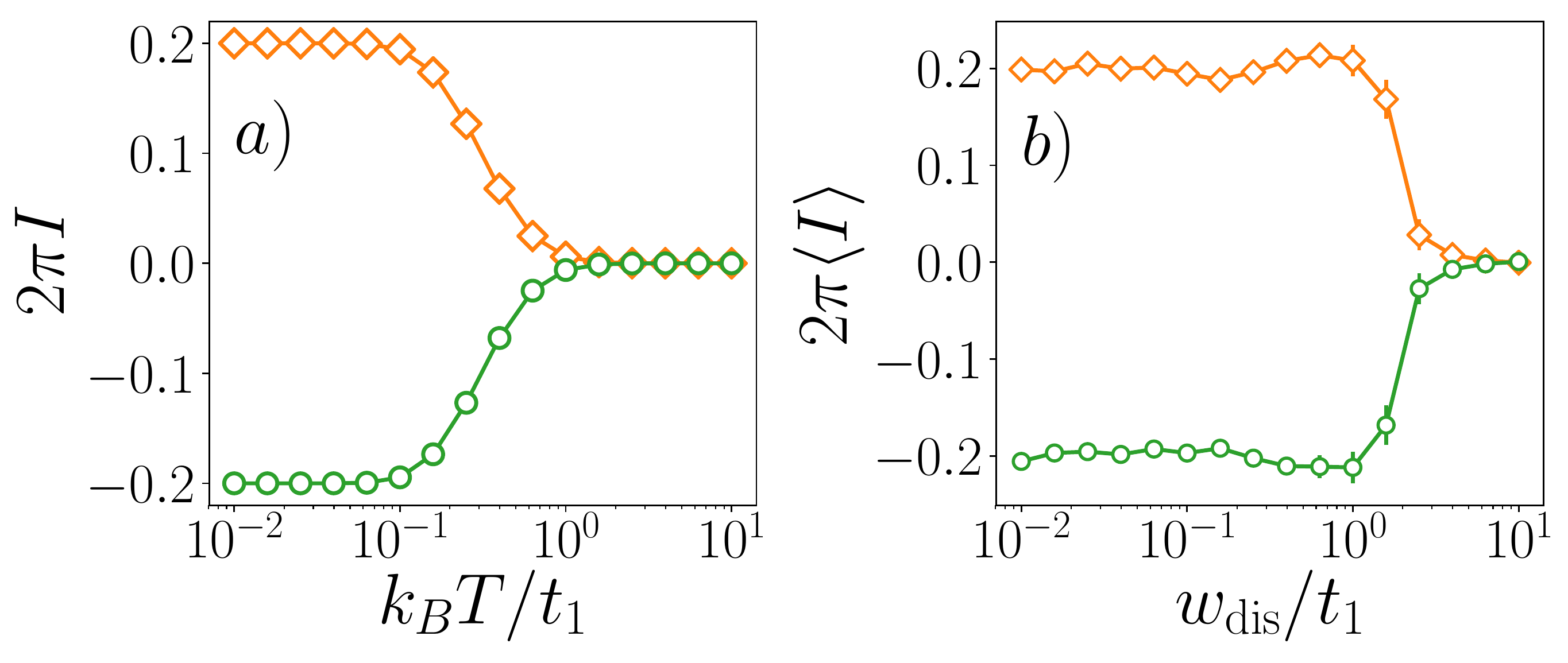}
	\caption[current]{Investigating the robustness of the bulk currents. We study a cylinder with a central stripe of increased potential, illustrated in Fig.~\ref{fig:middle-step-potn-currents}(a). Bulk currents are robust against (a) temperature and (b) disorder (here $T=0$). Data for (a) is presented for a cylinder with $L_x\times L_y=1000\times 30$, while (b) is presented for a cylinder $L_x\times L_y=10000\times 30$. Disorder is averaged over 100 samples and error bars on the mean $\langle I \rangle$ are only noticeable around the energy gap.}
	\label{fig:middle-step-potn-data}
\end{figure}

The topological nature of bulk currents manifests also in a resilience to thermal excitations and disorder. First, Fig.\ \ref{fig:middle-step-potn-data}(a) shows the same temperature dependency for bulk currents as we saw for the edge currents [Fig.~\ref{fig:global-mu-currents}(c)]. And second, Fig.\ \ref{fig:middle-step-potn-data}(b) confirms that bulk currents are largely insensitive to local random potentials $V(\boldsymbol{r}) = \delta,$ with $\delta$ drawn with uniform probability from the interval $\delta\in[-w_\text{dis},w_\text{dis}].$ Note that in both cases---temperature and disorder acting on edge or bulk currents---the topological currents can only survive a perturbation with a strength below the insulator gap, for otherwise the system may develop particle and hole excitations that partially cancel the currents.

\subsection{Currents and localised states}

\begin{figure}[t]
\centering
\includegraphics[width=1\columnwidth]{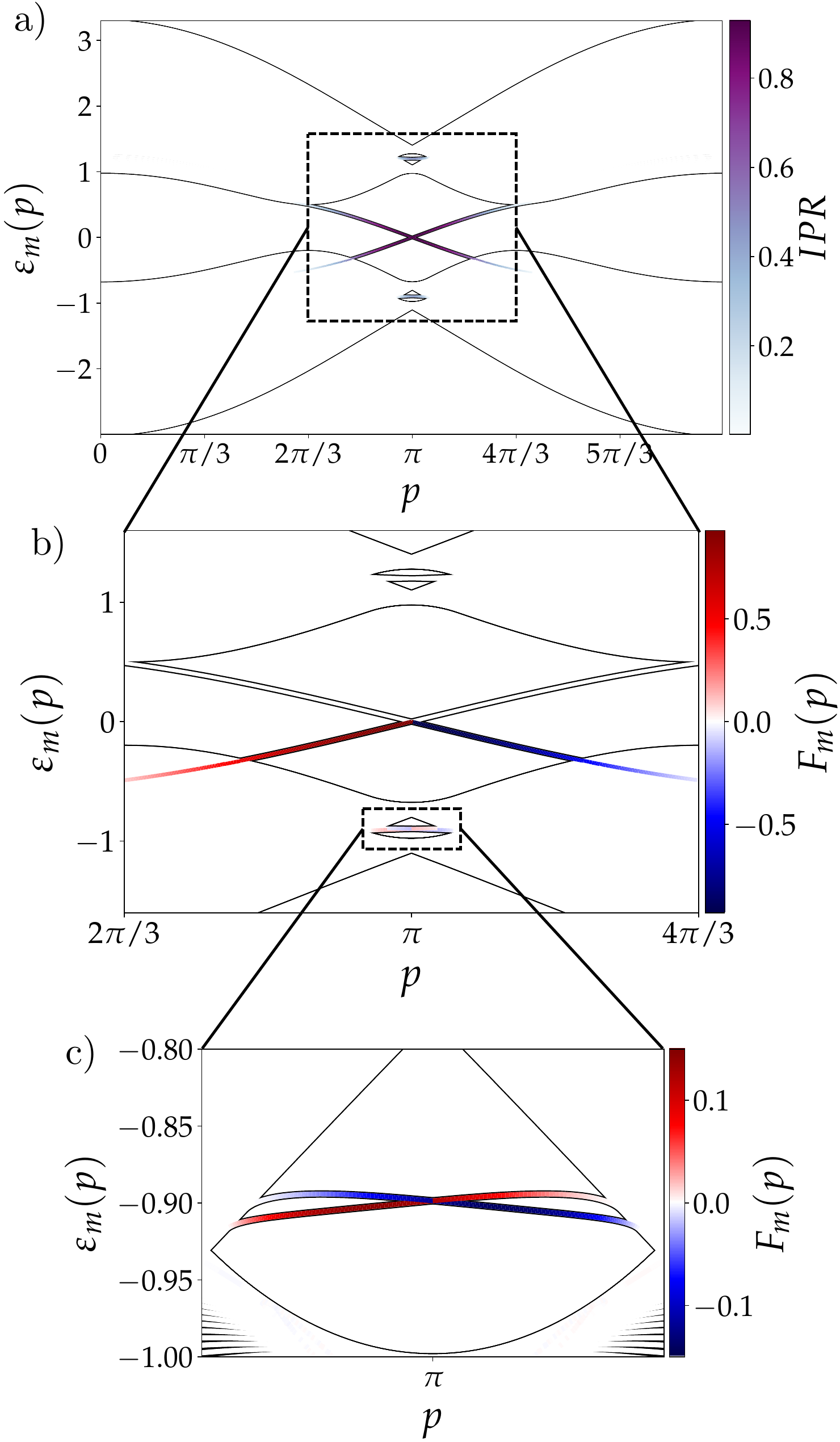}
\caption{The spectrum of the system showing localisation measures of the states. The setup is a cylinder with a central stripe of increased potential, as shown in Fig.~\ref{fig:middle-step-potn-currents}(a). (a) The IPR, Eqn.~\eqref{eq:inverse_pariticipation_ratio} is shown for all states in the system. Highlighting strongly localised edge states as well as localised bulk states. (b) The asymmetrical IPR $F_{n,p}$, Eqn.~\eqref{eq:asymmetrical_IPR}, plotted for states close to zero energy and momentum. (c) Zooming in on the states localised within the bulk, we can infer they give rise to bulk currents.}
\label{fig:IPR}
\end{figure}

It is well known that edge currents arise from states that are exponentially localised on the edge of the system. A natural question is whether the states participating in bulk currents share the same localisation properties. We characterize the localisation strength with the \textit{inverse participation ratio} (IPR). The $\text{IPR}_{m,p}$ for the $m$-th eigenstate $\psi_{m,p}$ with quasimomentum $p$ is the fourth moment of the wave function
\begin{equation}
\text{IPR}_{m,p} = \sum_y |\psi_{m,p}(y)|^4 \, .
\label{eq:inverse_pariticipation_ratio}
\end{equation}
States with a large value of IPR are strongly localised in space. We can obtain more information from the IPR by considering where states are localised in the system. To study this we define the asymmetrical IPR $F_{m,p}$
\begin{equation}
F_{m,p} = n_{T,\mu}\left( \varepsilon_{m}(p)\right)  \sum_{y} |\psi_{m,p}(y)|^4 \text{sign}\left(y-\frac{L_y}{2}\right).
\label{eq:asymmetrical_IPR}
\end{equation}
This assigns a positive value to states primarily localised in the lower half of the cylinder and a negative value to those localised to the upper half. It weights each state by its fermion occupation, $n_{T,\mu}\left( \varepsilon_{m}(p)\right)$, in order to indicate which states actually contribute to the particle current. At zero temperature and chemical potential  $T=\mu=0$ this will assign a zero value to all states above $\varepsilon = 0$.

We have investigated both quantities for a cylinder with a stripe potential, Eqn.~\eqref{eq:stripe}. Fig.~\ref{fig:IPR}(a) shows the IPR for a cylinder with a stripe potential,  confirming two distinct sets of localised states. First, there are highly localised gapless states that live on the boundaries of the cylinder. Considering these gapless states, in Fig.~\ref{fig:IPR}(b) we see the positive dispersing branch is localised in the lower half and the negative dispersing branch in the upper half. Since the potential is zero at the boundaries $V(0)=V(L_y)=0,$ these states are symmetrically filled and give a zero net edge current.

In the bulk, the potential stripe splits the bulk modes into two groups, one for each of the homogeneous local potential regions in the cylinder. This separation, best illustrated in Fig.~\ref{fig:IPR}(c), opens a window in the valence band of bulk states, with a mini-gap of size roughly \ $\simeq w$ for\ $t_2 \ll t_1$ [cf. Figs.~\ref{fig:IPR}(a-c)]. Within this small window, we find a new family of topological localised bulk states, connecting the two valence bands. As in the case of edge states, Fig.~\ref{fig:IPR}(c) shows a positively dispersing branch localised in the lower half and a negatively dispersing branch in the upper half. It is these states that give rise to the bulk currents, as we argue in the next Section. However, unlike edge states, which live in the insulator gap, these topological bulk states result from a continuous distortion of the original valence bands around $p\simeq \pi.$ Moreover, the strength of the current is not driven by changes in the occupation of states as localised bulk states are always filled. Instead, increasing the size of the potential step $w$ distorts the spectrum and widens the bulk energy windows, adding more states onto the fully filled localised bands shown in Fig.~\ref{fig:IPR}(c) and increasing the current it supports. This interpretation will be further confirmed by perturbation theory in the following section.

\section{Bulk currents from perturbation theory}
\label{sec:Origin-of-bulk-currents}

In this Section we show how localised cylinder modes can give rise to localised bulk currents. To that end, we analyse the momentum\ $p=\pi$ modes of the Haldane model on a cylinder. For these modes Hamiltonian~\eqref{eq:fourier_modes} becomes
\begin{equation}
\begin{split}
H(\pi) =&\ t_1\sum^{L_y}_{j=1} \left(b^\dagger_{j} a_{j+1} + \textrm{h.c.}\right) \\
&+ t_2 \sum_j \left(2i\ a^\dagger_{j} a_{j+1} -2i\ b^\dagger_{j} b_{j+1} + \textrm{h.c.}\right) \\
&+ \sum_j \left[ V_a(j) a^\dagger_{j} a_{j} +  V_b(j)  b^\dagger_{j} b_{j}\right]  .
\end{split}
\label{eq:pi_modes}
\end{equation}
It consists of a 1D SSH chain with perfectly dimerised first neighbour tunnelings\ $t_1$ inside an inhomogeneous potential\ $V_a,\ V_b$ and with complex second neighbour tunnelings\ $t_2$. We consider a single potential step as the simplest potential configuration that hosts localised modes in the bulk
\begin{equation}
V_a(j) = V_b(j) =
\left\{\begin{array}{ll}
-w,& j \leq L_y/2 \\
+w,& j > L_y/2
\end{array}\right. .
\label{eq:step_potential}
\end{equation}

\subsection{Analysis of the 1D SSH chain}

\begin{figure}[t]
	\centering
	\includegraphics[width=\linewidth]{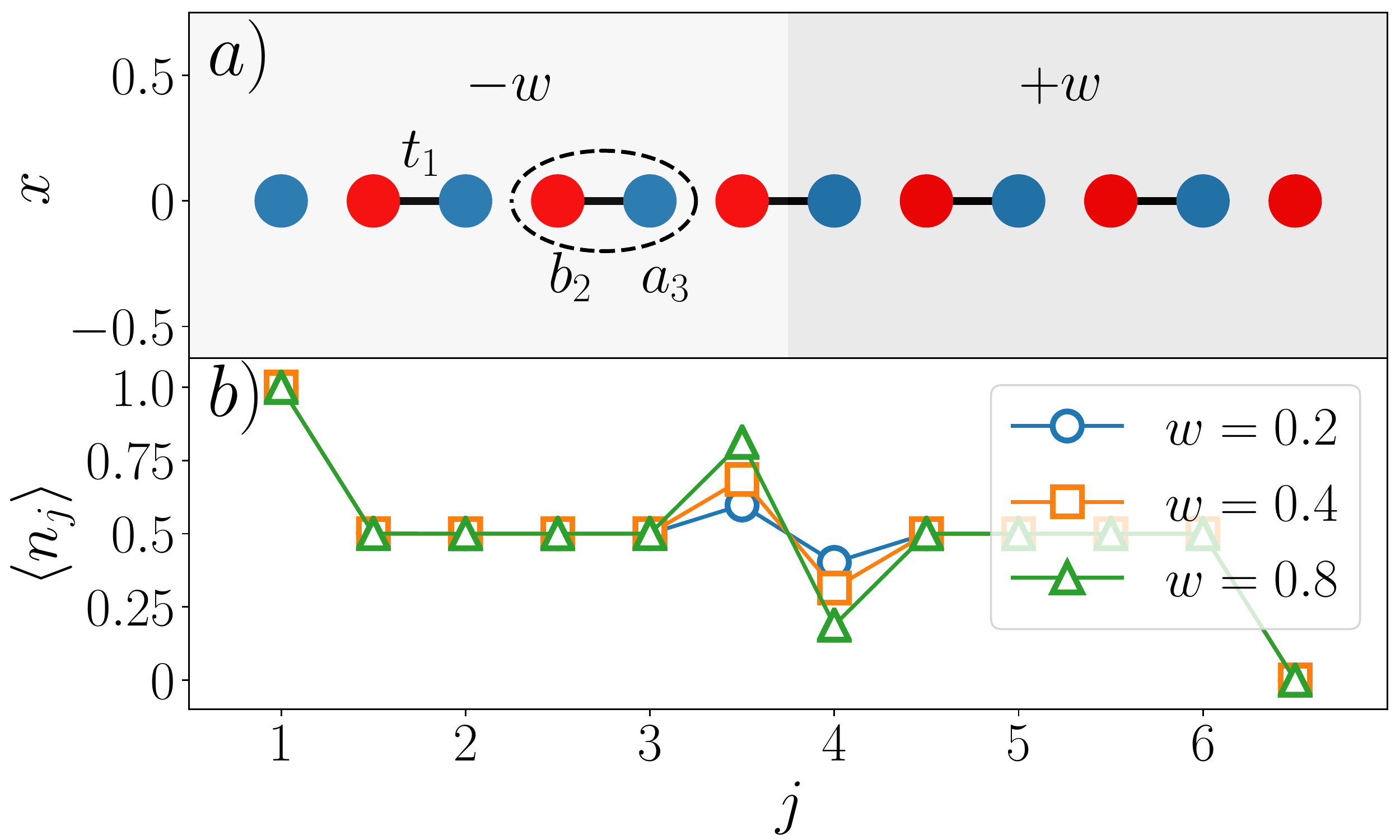}
	\caption{Population distribution of a simple SSH chain inside the step potential of Eqn.~\eqref{eq:step_potential}. (a) The\ $t_1$ coupling connects pairs of alternated sites, or dimers. An example of such is formed by the pair of sites\ $(b_2, a_3)$, shown inside the dashed circle. The two edge sites are disconnected from the rest of the chain. (b)~The population density is evenly distributed inside each dimer, except for the dimer at the potential step. To illustrate the population imbalance at the step we show\ $w=0.2,0.4,0.8$. The edge modes are occupied or unoccupied depending on whether they have negative or positive potential, respectively, with average population $0.5$ between them. Data is show for a chain of\ $L_y = 6$ at\ $T=0$.}
	\label{fig:ssh_analysis}
\end{figure}

We begin by studying the SSH model that appears when we only have\ $t_1$ couplings inside a potential step, with $t_2=0$. In this limit, all fermions are localised in dimers, which are pairs of sites\ $(b_{j-1},\ a_j)$ connected by a\ $t_1$ coupling, like the two sites inside the dashed circle in Fig.~\ref{fig:ssh_analysis}(a). There are also two localised edge modes, each formed by a single site. These two edge modes are decoupled from the $t_1$ terms of the Hamiltonian and have populations $n_1$ and $n_N$ subject to the constraint $n_1+n_N=1$ due to half filling. As all fermions are localised in dimers, there can be no correlations between neighbouring dimers, i.e.\ $\langle a^\dagger_{j} b_{j}\rangle = 0$.

The dimer modes in the bulk are degenerate and their population is evenly distributed between their two sites, $n_{b_{j-1}} = n_{a_{j}} = 1/2$, as shown in Fig~\ref{fig:ssh_analysis}(b). Because they are degenerate in energy they can rearrange themselves into delocalised modes along the regions with equal local potential [cf. Fig.~\ref{fig:IPR}(a)]. However, the dimer modes located at the potential step, which we will refer to as step modes, have a different local potential at each dimer site. Therefore they have a different energy than the rest of the bulk modes. Moreover, there is an imbalance in the population of their two sites of the order of the potential step,\ $w$, as we show in Fig.~\ref{fig:ssh_analysis}(b). Since the step modes are non-degenerate, they are always localised at the step for\ $t_2 \ll t_1$.

\subsection{Localised currents in the cylinder}

For small\ next-nearest neighbour couplings $t_2 \ll t_1$ we can study the Hamiltonian modes of Eqn.~\eqref{eq:pi_modes} using a perturbation theory analysis. The full mathematical derivation is carried out in Appendix~\ref{sec:perturbation_theory}. As the next-to-nearest neighbour coupling $t_2$ is non-zero the bulk modes are no longer localised inside dimers, but in triplets of neighbouring dimers. That is, if we define a single dimer mode localised at sites\ $(b_{j-1},\ a_j)$ as\ $|j\rangle$, then the modes for small\ $t_2\ll t_1$ are
\begin{equation}
\begin{split}
|j(t_2)\rangle = |j\rangle &+ \xi_{-1}|j-1\rangle + \xi_{+1}|j+1\rangle, \\ &\xi_{-1},\xi_{+1}\propto \frac{t_2}{t_1}.
\end{split}
\end{equation}
This small delocalisation of all bulk modes induces correlations between neighbouring dimer sites\ $\langle a^\dagger_{j} b_{j}\rangle$ that depend on the differences between the energies and populations of the two neighbouring dimer modes [cf. Eqn.~\eqref{eq:dimer_correlation}].

\begin{figure}[t]
	\centering
	\includegraphics[width=\linewidth]{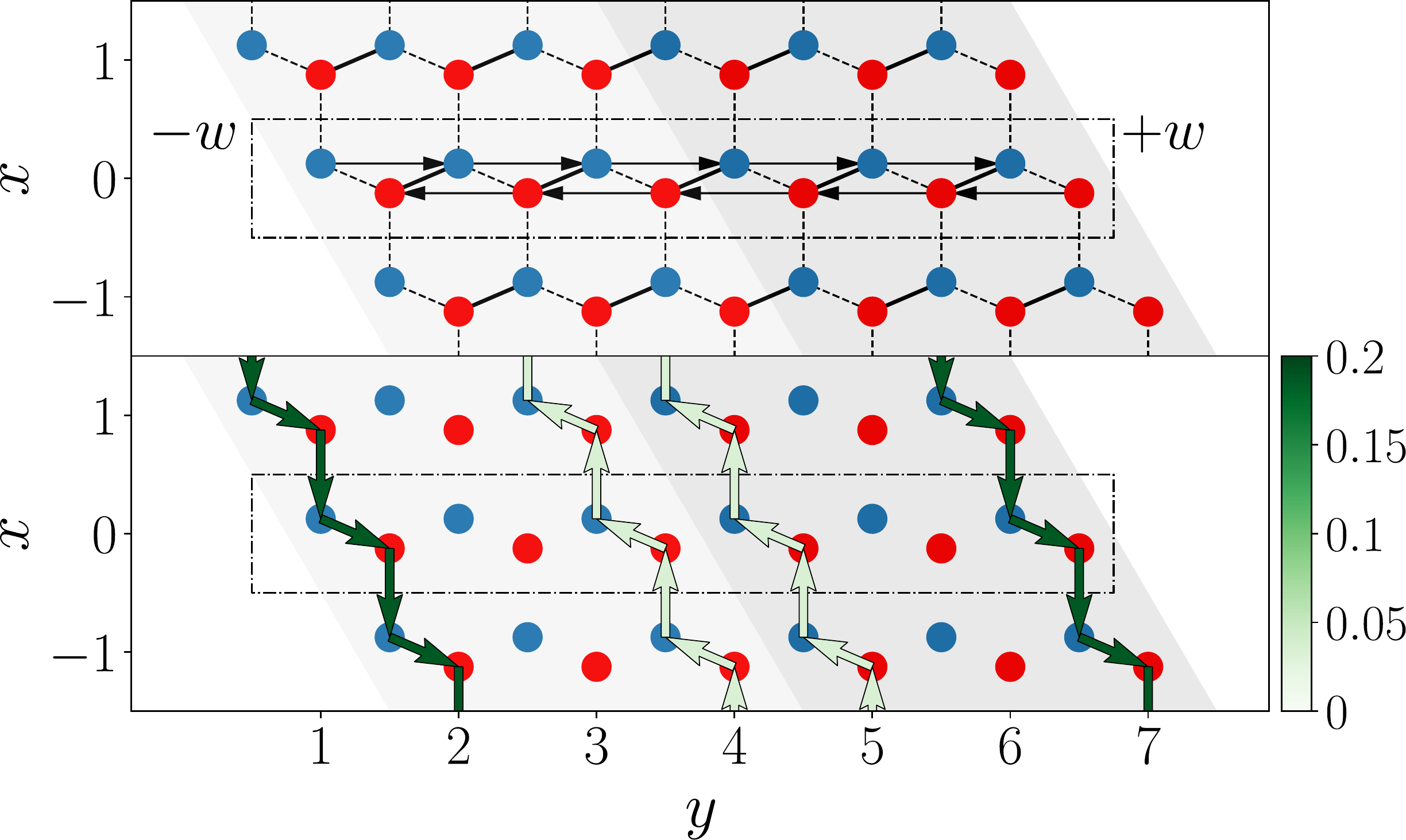}
	\caption{Currents in the Haldane cylinder with a potential step\ $w$, arising from the Fourier mode\ $p=\pi$. (a) Inside the dash-dotted rectangle we show a sample of periodic unit cell, for which we also show the couplings inside of it ($t_1$: dashed and continuous lines,\ $t_2$: arrow lines). The other continuous and dashed lines are the \ $t_1$ couplings of the whole Haldane model inside a cylinder. (b) Localised bulk currents and edge currents for a step potential\ $w=0.1$. Data presented for\ $L_y = 6,\ T=0$.}
	\label{fig:ladder_analysis}
\end{figure}

Coming back to the analysis of the Haldane Hamiltonian inside a cylinder, we can use Eqs.\ \eqref{eq:correlations} and \eqref{eq:two_point_currents} together with the dimer mode correlations\ $\langle a^\dagger_{j} b_{j}\rangle$ to study the two-point currents along a cylinder generated by the\ $p = \pi$ modes. In the homogeneous potential regions we have shown that all dimers are degenerate and have the same population imbalance. Therefore,\ $\langle a^\dagger_{j} b_{j}\rangle = 0$ and there are no currents inside those regions. However, the localised modes at the edges of the chain and at the potential step have non-zero correlations, resulting in localised currents at the cylinder boundaries and the potential step, respectively. For small couplings\ $t_2 \ll t_1$ and potential steps\ $w\ll t_1$, these localised currents are
\begin{equation}
\begin{split}
I_\text{bulk} &= - 4 \frac{t_2 w}{t_1} \\ 
I_\text{edge} &= 2 t_2 . 
\end{split}
\end{equation}
We show in Fig.~\ref{fig:ladder_analysis}(a) a sketch of the unit cell of the cylinder model and the couplings inside of it. Fig.~\ref{fig:ladder_analysis}(b) shows the localised currents around the cylinder boundaries and potential step due to the localisation of the edge and step bulk modes for momentum\ $p=\pi$.

\subsection{Dispersion relation}

The SSH chain with\ $t_2\ll t_1$ couplings explains how localised modes give rise to localised currents on the cylinder. However, we know [cf. Eqn.~\eqref{eqn:preden}] that a Chern Insulator only has currents when it has modes with nonzero dispersion relation. We can analyse the dispersion relation of the SSH chain modes by introducing a momentum offset\ $p = \pi + \delta p$ into the Hamiltonian modes of Eqn.~\eqref{eq:fourier_modes}. By treating this offset as a small perturbation potential\ $V_p$ [cf. Eqn.~\eqref{eq:offset_potential}], we can compute the dispersion relation of a mode\ $|\psi_j\rangle$ as
\begin{equation}
\begin{split}
\frac{dE_j}{dp}(\pi) &= \lim_{p\rightarrow 0} \frac{E_j\left(\pi + \delta p \right) - E_j\left(\pi\right)}{\delta p} \\
&= \frac{\langle \psi_j| V_p | \psi_j\rangle}{\delta p}.
\end{split}
\end{equation}
For small steps\ $w \ll t_1$ and couplings\ $t_2 \ll t_1$, we find for the localised step and edge modes
\begin{equation}
\frac{dE_\text{step}}{dp}(\pi) = \pm 6 \frac{t_2 w}{t_1},\quad \frac{dE_\text{edge}}{dp}(\pi) = \pm 6 t_2,
\end{equation}
with the sign depending on which cylinder boundary the mode is located at or whether the step mode has positive or negative energy. These results, although for a different potential, quantitatively agree with the dispersion relation of the modes shown in Fig.~\ref{fig:IPR}.

\section{Conclusions}
\label{sec:Conclusions}

Summing up, we have found that the bulk of a Chern insulator can exhibit topological localised bulk states when a gradient of a potential is present. These localised bulk states can support \textit{bulk currents.} They share an intrinsic robustness against temperature and local disorder with edge currents. However, while edge currents are confined at the boundaries of the material, bulk currents are defined by gradients of local potentials and can adopt any shape, extension and strength.

We qualitatively and quantitatively verified our numerical findings with perturbation theory. In this way we show that the bulk currents result from the disentangling of normal bulk modes caused by the presence of a potential gradient. Such a disentangling of modes in a topological material creates localised bulk modes that support the corresponding bulk currents. Our investigation of topological bulk currents opens the door to further studies in more complex models, such as fermions with spin\ \cite{kane2005quantum}, topological superconductors or topological phases with interactions.

Finally, our work suggest that localised bulk states can be used in the same applications as edge states, with greater versatility and without the need for complex sample fabrication and shaping. To this end, it is important to emphasize that, even though much of our quantitative study has been implemented with step-wise functions, this is not required for the implementation. Bulk currents may appear wherever there is a potential gradient, and a strong combined topological current may result from a weaker potential variation that spreads over a whole band of the material [cf. Fig.~\ref{fig:triangle-potn-currents}].

\acknowledgments

We would like to thank Sofyan Iblisdir for the initial suggestion to investigate the edge currents of the Haldane model and for exciting conversations throughout the project. We would like to thank Jos\'{e} Garr\'{e}-Rubio for exciting conversations. We would like to thank Jaakko Nissinen for pointing the IQHE connection given in \eqref{eqnIQHE}. This work was supported by the EPSRC grant EP/R020612/1, Spanish Projects PGC2018-094792-B-I00  (MCIU/AEI/FEDER, EU), PGC2018-094180-B-I00 (MCIU/AEI/FEDER, EU), FIS2015-63770-P (MINECO/FEDER, EU), CAM/FEDER Project No. S2018/TCS-4342 (QUITEMAD-CM) and CSIC Research Platform PTI-001. Statement of compliance with EPSRC policy framework on research data: This publication is theoretical work that does not require supporting research data.

\bibliographystyle{apsrev4-1}
\bibliography{references}

\appendix

\section{Perturbation theory}
\label{sec:perturbation_theory}
As mentioned in the main text, it is interesting to note that at momentum\ $p=\pi$ the model reduces to an approximate SSH model with complex second neighbor tunneling amplitudes\ $t_2$ that we can treat in a perturbative approach
\begin{equation}
\begin{split}
H = &\ H_0 + V_t \\
H_0 = &\ t_1\sum^{L_y}_{j=1} \left(b^\dagger_{j} a_{j+1} + \textrm{h.c.}\right) \\
&+ \sum_j \left[V_a(j) a^\dagger_{j} a_{j} + V_b(j) b^\dagger_{j} b_{j}\right] \\
V_t = &\ t_2 \sum_j \left(2i\ a^\dagger_{j} a_{j+1} -2i\ b^\dagger_{y} b_{j+1} + \textrm{h.c.}\right).
\label{eq:perturbed_ssh_model}
\end{split}
\end{equation}
We analyze the case of a potential step of Eqn.~\eqref{eq:step_potential} at zero temperature as the simplest one that shows both localized bulk and edge currents. The step is located between sites\ $b^\dagger_{s-1}$ and\ $a^\dagger_s$.

\subsection{SSH chain modes}
The eigenmodes of the unperturbed Hamiltonian\ $H_0$ are localized dimers of the form
\begin{equation}
\begin{split}
&|j, -\rangle_0 = \left(\alpha_j a^\dagger_j - \beta_{j-1} b^\dagger_{j-1} \right)|0\rangle \\
&|j, +\rangle_0 = \left(\beta_{j-1} a^\dagger_j + \alpha_j b^\dagger_{j-1} \right) |0\rangle.
\end{split}
\end{equation}
with\ $\alpha_j,\ \beta_j > 0$ and\ $|j,-\rangle$ the mode with lowest energy. Additionally, there are two modes located at the edges of the chain
\begin{equation}
|L\rangle_0 = a^\dagger_1 |0\rangle,\qquad |R\rangle_0 = b^\dagger_{L_y} |0\rangle.
\end{equation}

All dimer modes have\ $\alpha_j = \beta_{j-1} = 1/\sqrt{2}$ and, therefore, their population is evenly distributed between the two dimer sites, except for the dimer located at the potential step, which satisfies for small steps\ $w\ll t_1$
\begin{equation}
\begin{split}
\alpha_s &= \sqrt{\frac{r-w}{2r}} = \frac{1}{\sqrt{2}}\left(1- \frac{w}{2t_1} \right) + \mathcal{O}\left(\frac{w^2}{t^2_1}\right) \\
\beta_{s-1} &= \frac{t_1}{\sqrt{2r(r-w)}} = \frac{1}{\sqrt{2}}\left(1 + \frac{w}{2t_1} \right) + \mathcal{O}\left(\frac{w^2}{t^2_1}\right),
\end{split}
\label{eq:perturbed_amplitudes}
\end{equation}
with\ $r = \sqrt{t^2_1 + w^2}$.

The energy of the dimers at the step is\ $\varepsilon^0_{s\pm} = \pm r$, while the energy of the rest of the bulk modes is\ $\varepsilon^0_{j\pm} = \pm t_1 - w$ if they are located in a region with\ $-w$ homogeneous local potential and\ $\varepsilon^0_{j\pm} = \pm t_1 + w$ if they are located in a region with\ $+w$. Additionally, for this step potential, the edge modes have energies\ $\varepsilon^0_L = -w,\ \varepsilon^0_R = +w$. A summary of the energies of the simple SSH chain can be found in Table~\ref{tab:relevant_values_perturbation_theory}. Note that the bulk dimer modes that are not at the step are degenerated. Therefore, they can rearrange themselves into delocalized modes along the bulk, as shown in Fig.~\ref{fig:IPR}.

\subsection{First order perturbation expansion}

Using these simple modes of the unperturbed Hamiltonian we can show that the perturbation potential\ $V_t$ does not modify the energies of the edge and dimer modes at first order
\begin{equation}\label{key}
\varepsilon^1_{j,\phi} = \langle j,\phi |_0 V_t|j,\phi\rangle_0 = 0,\quad\forall j\in \left[1,L_y \right] ,\phi \in \left\lbrace +,-\right\rbrace ,
\end{equation}
because\ $V_t$ connects terms in different dimers. However, the modes of the Hamiltonian are modified by perturbation potential as
\begin{equation}
|j,\phi\rangle_1 = | j,\phi \rangle_0 + \sum_{k: \varepsilon^0_{k,\phi'} \neq \varepsilon^0_{j,\phi}} \frac{\langle k,\phi' |_0 V_t | j,\phi \rangle_0}{\varepsilon^0_{j,\phi} - \varepsilon^0_{k,\phi'}} |k,\phi'\rangle_0.
\end{equation}
Because\ $V_t$ connects only second neighbor sites, a perturbed dimer mode\ $| j,\phi \rangle_0$ consists of a linear combination of itself and its neighboring ones\ $| j\pm 1,\phi' \rangle_0 $.

\subsection{Correlations inside the chain}
\label{subsec:appendix_correlations}

Let's compute the contribution of a single perturbed eigenstate to correlation terms of the type\ $\langle a^\dagger_j b_j\rangle$ that join neighboring dimers. We define
\begin{equation}
\begin{split}
\theta_{j\phi} &= \langle j,\phi|_1 a^\dagger_j b_j|j,\phi\rangle_1 \\
\varphi_{j\phi} &= \langle j,\phi|_1 a^\dagger_{j-1} b_{j-1}|j,\phi\rangle_1.
\end{split}
\end{equation}
For the lowest energy modes\ $\phi=-$ we find
\begin{equation}\begin{split}
\theta_{j-} =& \ t_2 \langle j,-|_0a^\dagger_j b_j|j+1,-\rangle_0 \frac{\langle j+1,-|_0 V_t |j,-\rangle_0}{\varepsilon^0_{j-}-\varepsilon^0_{j+1,-}}\\
&+ t_2 \langle j,-|_0a^\dagger_j b_j|j+1,+\rangle_0 \frac{\langle j+1,+|_0 V_t |j,-\rangle_0}{\varepsilon^0_{j-}-\varepsilon^0_{j+1,+}} \\
=&\ 2it_2 \alpha_j\beta_j \frac{\alpha_j\alpha_{j+1} - \beta_{j-1}\beta_j}{\varepsilon^0_{j-}-\varepsilon^0_{j+1,-}} \\
&- 2it_2\ \alpha_j \alpha_{j+1} \frac{\alpha_j\beta_j + \alpha_{j+1}\beta_{j-1}}{\varepsilon^0_{j-}-\varepsilon^0_{j+1,+}},
\end{split}
\end{equation}
\begin{equation}
\begin{split}
\varphi_{j-} =& \ t_2 \langle j-1,-|_0a^\dagger_{j-1} b_{j-1}|j,-\rangle_0 \frac{\langle j-1,-|_0 V_t |j,-\rangle^*_0}{\varepsilon^0_{j-}-\varepsilon^0_{j-1,-}} \\
&+ t_2 \langle j-1,+|_0a^\dagger_{j-1} b_{j-1}|j,-\rangle_0 \frac{\langle j-1,+|_0 V_t |j,-\rangle^*_0}{\varepsilon^0_{j-}-\varepsilon^0_{j-1,+}} \\
=&\ 2it_2 \alpha_{j-1}\beta_{j-1} \frac{\alpha_{j-1}\alpha_j - \beta_{j-2}\beta_{j-1}}{\varepsilon^0_{j-}-\varepsilon^0_{j-1,-}} \\
&+ 2it_2\ \beta_{j-2} \beta_{j-1} \frac{\beta_{j-2}\alpha_j + \alpha_{j-1}\beta_{j-1}}{\varepsilon^0_{j-}-\varepsilon^0_{j-1,+}},
\end{split}
\end{equation}
where the fractional terms don't contribute if the denominator is zero. In Table\ \ref{tab:relevant_values_perturbation_theory} we present relevant values of\ $\theta_{j-},\ \varphi_{j-}$ for the perturbed SSH model.

With this we can compute the expectation value of the correlations between dimers
\begin{equation}
\langle a^\dagger_j b_j\rangle = \sum_{\phi=+,-} \theta_{j\phi} n\left( \varepsilon_{j\phi}\right)  + \varphi_{j+1,\phi} n\left( \varepsilon_{j+1,\phi}\right)  ,
\end{equation}
with\ $n\left( \varepsilon_{j\phi}\right)$ the occupation of dimer mode\ $j\phi$, which follows Fermi-Dirac statistics. Therefore the correlations between dimers are
\begin{equation}\begin{split}
\langle a^\dagger_j b_j\rangle =&\ \theta_{j-} + \varphi_{j+1,-} \\
=&\ 2it_2 \left(\beta_{j-1}\alpha_{j+1} + \alpha_j\beta_j\right) \\
&\times \left(\frac{\beta_{j-1} \beta_j}{\varepsilon^0_{j+1,-}-\varepsilon^0_{j+}} - \frac{\alpha_j \alpha_{j+1}}{\varepsilon^0_{j-}-\varepsilon^0_{j+1,+}} \right) .
\label{eq:dimer_correlation}
\end{split}
\end{equation}
This correlation is zero if the two neighboring dimer modes have the same energies\ $\varepsilon^0_{j+1,-} = \varepsilon^0_{j-},\ \varepsilon^0_{j+1,+} = \varepsilon^0_{j+}$ and a balanced population density distribution\ $\beta_{j-1} \beta_j = \alpha_j \alpha_{j+1}$. This is the case for all correlations between dimer modes in the same potential region. However, the potential step modifies the energy and population distribution of the dimer mode inside it, thus that mode has a non zero correlation between itself and its two neighboring dimer modes. The same happens for the edge modes, which are intrinsically different from the bulk modes and a non zero correlation appears between them and their respective neighbor dimer modes.

\subsection{Two-point currents in the cylinder}
These correlations give us access to the two point currents between sites in the lattice. We can use Eqs.\ \eqref{eq:correlations} and \eqref{eq:two_point_currents} together with the perturbed dimer modes to compute the contribution of the\ $p=\pi$ momentum modes to the localized currents along the cylinder. At the potential step we find two localized currents, one at each side of the step, of the form
\begin{equation}
\begin{split}
I_{a_{s-1},b_{s-1}} &= -2 \frac{t_2 w}{t_1} + \mathcal{O}\left(t_2\frac{w^2}{t^2_1}\right) \\
I_{a_{s},b_{s}} &= -2 \frac{t_2 w}{t_1} + \mathcal{O}\left(t_2\frac{w^2}{t^2_1}\right).
\end{split}
\end{equation}
At the cylinder boundaries we find
\begin{equation}
I_{a_1,b_1} = 2t_2,\quad I_{a_{L_y},b_{L_y}} = 2t_2 .
\end{equation}
However, at the bulk of the cylinder, the correlations between dimer sites are zero, so there are no currents along its bulk.

\subsection{Dispersion relation at\ $p=\pi$}
\label{subsec:dispersion_relation}

The dispersion relation tells us how much the energy of a single mode changes when we vary the momentum
\begin{equation}
\frac{d\varepsilon_j}{dp}(p) = \lim_{p\rightarrow 0} \frac{\varepsilon\left(p + \delta p \right) - \varepsilon\left(p\right)}{\delta p}  .
\end{equation}
We can model this energy change at momentum\ $p = \pi$ by introducing an offset potential term\ $V_p$ in the perturbed SSH Hamiltonian that simulates and offset in the momentum\ $p = \pi + \delta p$. Then, analyzing how much this potential modifies the mode energies at\ $p = \pi$, we can write the dispersion relation of the SSH model as
\begin{equation}
\frac{d\varepsilon_j}{dp}(p=\pi) = \frac{\langle j,-|_1 V_p\left(\delta p\right) | j,-\rangle_1}{\delta p},
\label{eq:dispersion_relation}
\end{equation}
with the offset perturbation potential
\begin{equation}
\begin{split}
V_p = &\ \delta p\ t_1\sum^{L_y}_j \left(-ia^\dagger_j b_j + \textrm{h.c.}\right) + 2 \delta p\ t_2 \sum_j \left( a^\dagger_j a_j - b^\dagger_j b_j\right)  \\
&+ \delta p\ t_2 \sum_j \left(- a^\dagger_j a_{j+1}  + b^\dagger_j b_{j+1} + \textrm{h.c.}\right).
\label{eq:offset_potential}
\end{split}
\end{equation}
We define for each mode\ $|j,\phi\rangle_1$
\begin{equation}
\begin{split}
\chi_j =& \langle j,\phi|_1 \left( a^\dagger_j a_j - b^\dagger_{j-1}b_{j-1}\right) |j,\phi\rangle_1 \\
\sigma^a_j =& \langle j,\phi|_1 a^\dagger_j a_{j+1}|j,\phi\rangle_1 \\
\sigma^b_j =& \langle j,\phi|_1 b^\dagger_{j-1} b_j|j,\phi\rangle_1 \\
\tau^a_j =& \langle j,\phi|_1 a^\dagger_{j-1} a_j|j,\phi\rangle_1 \\
\tau^b_j =& \langle j,\phi|_1 b^\dagger_{j-2} b_{j-1}|j,\phi\rangle_1,
\end{split}
\end{equation}
then, the dispersion relation\ \eqref{eq:dispersion_relation} of a mode\ $|j,\phi\rangle_1$ is
\begin{equation}
\begin{split}
\frac{d\varepsilon_{j\phi}}{dp} = &\ 2 t_1\ \textrm{Im}\left\lbrace \theta_{j\phi} + \varphi_{j\phi} \right\rbrace + 2  t_2 \chi_{j\phi} \\
&+ 2  t_2\ \textrm{Re}\left\lbrace -\sigma^a_{j\phi} - \tau^a_{j\phi} + \sigma^b_{j\phi} + \tau^b_{j\phi} \right\rbrace
\end{split}
\end{equation}
To first order in\ $t_2$, the elements\ $\theta, \varphi, \sigma, \tau$ are purely imaginary, so the dispersion relation is
\begin{equation}
\frac{d\varepsilon_{j\phi}}{dp} = \ 2 t_1\ \textrm{Im}\left\lbrace \theta_{j\phi} + \varphi_{j\phi} \right\rbrace + 2 t_2 \chi_{j\phi}.
\end{equation}
In Table\ \ref{tab:relevant_values_perturbation_theory} we present relevant values for the dispersion relation of the negative energy modes. We find that the dispersion relation of the mode localized at the potential step is
\begin{equation}
\frac{d\varepsilon_{s-}}{dp} = - 6 \frac{w t_2}{t_1} + \mathcal{O}\left(t_2\frac{w^2}{t^2_1}\right),
\end{equation}
while the edge modes have
\begin{equation}
\frac{d\varepsilon_L}{dp} = 6 t_2,\quad \frac{d\varepsilon_{R}}{dp} = - 6 t_2.
\end{equation}
The bulk modes all have zero dispersion relation except for the localized modes neighboring the edge and step potential sites
\begin{equation}
\begin{split}
\frac{d\varepsilon_{2-}}{dp} &= -\frac{d\varepsilon_{s-1,-}}{dp} \\
&= \frac{d\varepsilon_{s+1,-}}{dp} = -\frac{d\varepsilon_{L_y,-}}{dp} = -t_2.
\end{split}
\end{equation}

\begin{table}[hbtp]
	\caption{Relevant values of\ $\theta, \varphi, \chi, \varepsilon$ to first order in\ $w$ for negative energy modes. $L\ (j=1)$ and $R\ (j=L_y)$ are the left and right edge modes, respectively, and the mode located at the potential step has\ $j=s$.}
	\begin{tabular}{| c | c | c | c | c |}
		\hline
		j & $\varphi_{j-}$ & $\theta_{j-}$ & $\chi_{j-}$ & $\varepsilon_{j-}$  \\ \hline
		$L$ & $0$ & $2i\frac{t_2}{t_1}$ & $1$ & $-w$ \\ \hline
		$2$ & $-i\frac{t_2}{t_1}$ & $i\frac{t_2}{2t_1}$ & $0$ & $-t_1-w$ \\ \hline
		$2<j<s-1$ & $-i\frac{t_2}{2t_1}$ & $i\frac{t_2}{2t_1}$ & $0$ & $-t_1-w$ \\ \hline
		$s-1$ & $-i\frac{t_2}{2t_1}$ & $i\frac{t_2}{t_1}$ & $0$ & $-t_1-w$ \\ \hline
		$s$ & $-i\frac{t_2\left( 1+w\right)}{t_1}$ & $i\frac{t_2\left( 1-w\right)}{t_1}$ & $-\frac{w}{t_1}$ & $-\sqrt{t^2_1+w^2}$ \\ \hline
		$s+1$ & $-i\frac{t_2}{t_1}$ & $i\frac{t_2}{2t_1}$ & $0$ & $-t_1+w$ \\ \hline
		$s+1<j<L_y$ & $-i\frac{t_2}{2t_1}$ & $i\frac{t_2}{2t_1}$ & $0$ & $-t_1+w$ \\ \hline
		$L_y$ & $-i\frac{t_2}{2t_1}$ & $i\frac{t_2}{t_1}$ & $0$ & $-t_1+w$ \\ \hline
		$R$ & $-2i\frac{t_2}{t_1}$ & $0$ & $-1$ & $+w$ \\ \hline
	\end{tabular}
	\label{tab:relevant_values_perturbation_theory}
\end{table}

\end{document}